\documentstyle[preprint,aps,epsf,floats]{revtex}

\newcommand{\sect}[1]{\section{#1}\setcounter{equation}{0}}

\newcommand{\be}{\begin{equation}}
\newcommand{\ee}{\end{equation}}
\newcommand{\bea}{\begin{eqnarray}}
\newcommand{\eea}{\end{eqnarray}}
\newcommand{\p}{\partial}
\newcommand{\vp}{\varphi}

\begin{document}
\preprint{\tighten \vbox{\hbox{hep-th/0004206} }}
\title{Self-tuning in an Outgoing Brane Wave Model}
\author{Gary T. Horowitz$^{1},$ Ian Low$^{2,3},$ and A.\ Zee$^{1,2}$}
\address{\vspace{0.5cm}$^{1}$Department of Physics,\\
University of California, Santa Barbara, CA 93106 \\
$^{2}$Institute for Theoretical Physics,\\
University of California, Santa Barbara, CA 93106 \\
$^{3}$Department of Physics,\\
Carnegie Mellon University, Pittsburgh, PA 15213}
\maketitle

{\tighten

\begin{abstract}

We introduce a new brane-world model in which the bulk
solution consists of outgoing plane waves. This is
an exact solution to string theory with no naked singularities.
The recently discussed self-tuning mechanism to
cancel the cosmological constant on a brane is naturally incorporated.
We show that even if the
vacuum energy on the brane changes, e.g. due to a phase transition,
the brane geometry remains insensitive to the local vacuum energy. We also
consider the static self-tuning branes introduced earlier, and find
a solution in which the brane geometry starts to contract when the vacuum
energy on the brane changes.

\end{abstract}

}



\newpage

\sect{Introduction}

The idea that our universe is embedded in a higher dimensional world has
received a great deal of renewed attention over the last 
two years\cite{n1,RS1}. 
Of particular interest is the fact that this opens up new approaches to 
solving the cosmological constant problem (see, e.g., 
\cite{BVV,KSS,HDKS})  though related ideas were 
pursued some time ago\cite{old,ant,vold}.
In particular, \cite{KSS,HDKS} show that
in a five dimensional theory of gravity coupled to a scalar field, the
brane geometry can remain flat, independent of the value of the cosmological
constant on the brane. This is known as ``self-tuning". 
There has been extensive discussion of this idea for static branes
 \cite{cosmo}. However,
a natural question to ask is whether the same self-tuning feature will
remain in a dynamical context where the vacuum energy on the brane 
changes, e.g., due to a phase transition. 

Dynamical discussions of our universe as a brane suffer from the awkward feature
that evolution requires not just initial data on the brane, but also initial
data for the bulk fields as well. Thus, there are a large number of degrees
of freedom that we do not have direct access to. The simplest, and most
natural assumption is that the these bulk degrees of freedom
simply respond to motion on the brane. In other words, there are no
incoming bulk waves, but only outgoing waves.
We construct and study this ``outgoing  brane wave model".
It has a number of desirable features: (1) The metric in the bulk is an
exact plane wave. Since the curvature is null, all higher order corrections
to Einstein's equation constructed from higher powers of the Riemann tensor
automatically vanish. Hence, these metrics are exact solutions to string
theory to all order in $\alpha'$ \cite{Ama,HS}. 
(2) The geometry on the brane is not
flat, but is typically given by an expanding Robertson-Walker 
cosmology\footnote{For other discussions of brane-world cosmology 
see \cite{cosmol}.}.
However, the rate of expansion is independent of the value of the cosmological
constant, so the self-tuning feature is incorporated. This continues to hold
even when the vacuum energy on the brane changes due to a phase
transition. (3) There are no
naked singularities. Instead, there are null singularities in the past
which extend from the big bang on the brane. (4) As mentioned above,
this is a deterministic model in which the bulk geometry is 
determined solely by the brane.

Even though one does not
need to specify boundary conditions at the null singularity,
it still cuts off space in the fifth dimension. So the proper distance
to the singularity on appropriate spacelike surfaces is finite, and the local
four dimensional Newton's constant will be nonzero. However, it is also
time dependent. This is a problem with this model which
is currently under investigation. Perhaps by adding a bulk cosmological 
constant
one could localize  gravity near  the brane \cite{RS1} and stabilize the
effective Newton's constant.

We also find  exact, time dependent solutions to the bulk
equations which are more general than plane waves. These solutions
can be used
to study a phase transition in which the vacuum energy changes
on an initially static, Poincare invariant brane.  We 
find a solution
in which the brane becomes time dependent after the transition.
Nevertheless, it remains an open question whether or not
 the brane becomes time dependent in all solutions with an
initially static brane.

After discussing the basic equations of motion in the next section,
we introduce our outgoing  brane wave model in section III and study its
properties. We then consider more general time dependent solutions
in section IV and use them to study phase transitions on initially 
static branes
in section V.

\sect{Equations of Motion}

Following \cite{KSS,HDKS}, we take the low energy effective action to be 
\begin{equation}
\label{action}
S=\int d^{5}x\sqrt{-G}\left[R-\frac{4}{3}(\nabla \varphi )^{2}\right]
+\int d^{4}x\sqrt{-g}\;e^{b\,\varphi}\,{\cal L}_{4D}  
\end{equation}
Gravity and the dilaton field $\varphi $ live in a 5-dimensional world (with
coordinates $t,$ $x^i,$ and $y)$ and are coupled to a thin 4-dimensional
brane whose position is taken to be at $y=0.$ The metric $g_{\mu \nu
}= \delta _{\mu }^{M}\,\delta _{\nu }^{N}\,G_{MN}$ is the 4-dimensional
metric induced on the brane, and ${\cal L}_{4D}$ denotes the Lagrangian of the
4-dimensional world. (Here $M,N=0,1,2,3,5$;  $\mu ,\nu =0,1,2,3$; and
we have set $16\pi G_5=1$.) We take
$b$ as a free phenomenological parameter which ultimately may be determined
by string theory. We will assume the geometry on the brane is homogeneous
and isotropic, so the stress energy tensor must take the form of a perfect
fluid:
\begin{eqnarray}
\label{stress}
T_{\;\;\nu}^{\mu }&=& e^{b\,\varphi}\,\mbox{diag}(-\rho _{T},P_{T},P_{T},P_{T})
  \nonumber\\
&=& -e^{b\,\varphi}\,V\,\delta _{\;\;\nu }^{\mu }
 + e^{b\,\varphi}\,\mbox{diag}(-\rho ,P,P,P)
\end{eqnarray}
where we have set $\rho_{T}=\rho+V $ and $P_{T}=P-V.$
With this form we describe a 4-dimensional world containing a vacuum energy $%
V$ and a perfect fluid with energy density $\rho$ and pressure $P$. 
We will
treat $V,$ $\rho ,$ and $P$ as phenomenological functions of $t.$

Einstein's equations read
\begin{equation}
  \label{eins}
R_{MN}-\frac{1}{2}G_{MN}R =
\frac{4}{3}\left[\nabla _{M}\varphi \nabla _{N}\varphi -%
\frac{1}{2} G_{MN}(\nabla \varphi )^{2}\right]+\frac{1}{2}\sqrt{\frac{g}G}\,T_{\mu
\nu }\,\delta _{M}^{\mu }\,\delta _{N}^{\nu }\,\delta (y)
\end{equation}
with $g$ and $G$ the determinant of $g_{\mu \nu }$ and $G_{MN}$
respectively. The equation of motion for the scalar field has a source
involving the the four dimensional Lagrangian
${\cal L}_{4D}$. The Lagrangian for a  perfect fluid 
(with pressure determined in terms of the 
energy density) is simply ${\cal L} = -\rho$ \cite{HE}. So if
we model the matter on the brane by a vacuum energy and perfect fluid,
we have ${\cal L}_{4D} = -(V+\rho)$. The dilaton equation is thus
\begin{equation}
\frac{8}{3}\nabla ^{2}\varphi =\sqrt{\frac{g}{G}}\,b\,e^{b\,\varphi
}(V+\rho )\, \delta (y) \label{scalar}
\end{equation}

We wish to consider solutions depending on both $t$ and $y$, which are 
homogeneous and isotropic
in the three transverse directions. It will be convenient to adopt 
``conformal gauge" in the $t,y$ directions, and write the metric in the form
\begin{eqnarray}
ds^{2}
&=&e^{2A(t,y)}(-dt^{2}+dy^{2})+e^{2B(t,y)}(dx_{1}^{2}+dx_{2}^{2}+dx_{3}^{2})
\label{metric} \nonumber\\
&=&-e^{2A(u,v)}\,du\,dv+e^{2B(u,v)}(dx_{1}^{2}+dx_{2}^{2}+dx_{3}^{2})
\end{eqnarray}
where as usual $u\equiv t-y$ and $v\equiv t+y.$

The easiest way to 
solve Einstein's equation (\ref{eins}) and the dilaton equation 
(\ref{scalar}) is to first solve them in the bulk,
 that is, away from the brane at $y=0.$ We will
solve these equations for $A(t,y),$ $B(t,y),$ and $\varphi (t,y)$ for $%
y>0 $ and $y<0$ separately and then match the solutions across the brane. We
find it convenient to use the $(u,v)$ light cone coordinates to solve the
equations in the bulk, and the $(t,y)$ coordinates to do the matching across
the brane.

Using the metric (\ref{metric})
we obtain the following bulk equations of motion 
\begin{eqnarray}
B_{,uv}+3B_{,u}\,B_{,v}&=&0  \label{153}\\
2\,\varphi _{,uv}+3\,(B_{,u}\,\varphi _{,v}+B_{,v}\,\varphi _{,u})&=&0 
\label{dil3}\\
A_{,u} B_{,u}-\frac{1}{2}(B_{,u}{}^2+B_{,uu})-\frac{2}{9}\varphi_{,u}{}^{2}
&=&0  \label{tt}\label{uu}\\
A_{,v}B_{,v}-\frac{1}{2}(B_{,v}{}^2+B_{,vv})-\frac{2}{9}\varphi_{,v}{}^{2}
&=&0  \label{553}\\
\label{ii15}
2\,\varphi _{,u}\,\varphi _{,v}+3(A_{,uv}+2B_{,uv} + 3B_{,u}\,B_{,v})&=&0  
\end{eqnarray}
Here (\ref{153}) is the $uv$ component of Einstein's equation. It is independent
of $\vp$ since the dilaton is essentially a two dimensional scalar
field and the $uv$ component of the stress energy tensor is like the trace of
the two dimensional stress tensor which vanishes. 
 (\ref{dil3}) is the dilaton equation of motion. Both of these first two
 equations are independent of $A$ essentially because the
 two dimensional wave equation is conformally invariant. (\ref{tt}) is the
$uu$ component of the Einstein equation, 
(\ref{553}) is the $vv$ component,  
and (\ref{ii15}) the $ii$ component. This last equation is a consequence
of the first four by virtue of the Bianchi identity.

The jump conditions for the metric at the brane
can be obtained from the usual thin
shell conditions of general relativity\footnote{They can also be obtained
by writing out (\ref{eins}) and integrating across $y=0$ at fixed $t$.}
\cite{IS}. 
The unit outward normal to the
surface $y=0$ is $n=  e^{-A} \p/\p y$. The extrinsic curvature $K_{\mu\nu} =
\nabla_\mu n_\nu$ is
\be 
K_{tt} = -e^A A_{,y} \qquad K_{ij} =  e^{2B-A} B_{,y}\ \delta_{ij}
\ee
and its trace is $K= {K_\mu}^\mu = e^{-A}(A_{,y} + 3B_{,y})$.
In general, if $K_{\mu\nu}]$ is the jump in the extrinsic curvature
across the surface $y=0$, the stress tensor on the brane is given by
\be
T_{\mu\nu} = -{1\over 8\pi G_5} \left(K_{\mu\nu}] - g_{\mu\nu} K]\right)
\ee
(For any function $F(t,y)$, we use the notation 
$\left.F\right]=F(t,0^{+})-F(t,0^{-})$
and $\left.F\right|=F(t,0)$.) 
Using the fact that we have set $16\pi G_5 =1$, and the form of the
stress tensor (\ref{stress}), we obtain the matching conditions
\begin{eqnarray}
-6 \,\left.\frac{\partial B}{\partial y}\right]
&=& \left.e^{A+b\,\varphi }\right|\rho_T =
 \left.e^{A+b\,\varphi }\right| (V+\rho ),  \label{Bmatch}\\
6 \, \left.\frac{\partial A}{\partial y}\right]
&=&\left.e^{A+b\,\varphi }\right|(2\rho _{T}+3P_{T})   \label{Amatch}\\
&=& \left.e^{A+b\varphi }\right|
(2\rho +3P -V) \nonumber \\ 
\frac{8}{3}\,\left.\frac{\partial \varphi}{\partial y}\right]
& =& b\,\left.e^{A+b\,\varphi }\right|(V+\rho ).
\label{phimatch} 
\end{eqnarray}
where the matching condition for the dilaton is obtained by
integrating  (\ref{scalar}) at fixed $t$ across $y=0$.

It is worth emphasizing that the matching conditions
have to be satisfied at any instant in $t$. In other words,
 both sides in the matching conditions (\ref{Bmatch}),
 (\ref{Amatch}) and (\ref{phimatch}) 
are functions of $t$, as opposed to be just constants in the static case.
Comparing (\ref{phimatch})
and (\ref{Bmatch}) we find
\begin{equation}
\left.\vp_{,y}\right] = -{9\over 4}b\,\left.B_{,y}\right]
\label{b}
\end{equation}
which is a rather strong constraint since $b$ is a constant.
In many of our solutions this condition forces $\varphi $ to be proportional
to $B.$

Note that in the 4-dimensional universe on the brane, the metric is given by
\begin{equation}
ds^{2}=- e^{2A(t,0)}\,dt^{2}+e^{2B(t,0)}(dx_{1}^{2}+dx_{2}^{2}+dx_{3}^{2})
\end{equation}
which can be written in the standard form of a Robertson-Walker
universe
\begin{equation}
ds^{2}=-d\tau ^{2}+a(\tau )^{2}(dx_{1}^{2}+dx_{2}^{2}+dx_{3}^{2})  \label{rw}
\end{equation}
with
\begin{eqnarray}
 \tau &=& \int dt\, e^{A(t,0)}, \\
 a(\tau )&=& e^{B(t(\tau ),0)}.
\end{eqnarray}

\sect{Outgoing  brane wave model}

There is a simple class of solutions to the bulk equations (\ref{153}) --
(\ref{ii15})
which consists of  
taking $A,$ $B,$ and $\varphi $ to
depend only on $u$.
The five field equations then collapse to just one
\begin{equation}
A_{,u} B_{,u}-\frac{1}{2}(B_{,u}{}^2+B_{,uu})-\frac{2}{9}\varphi_{,u}{}^{2}
=0  \label{Aeq}
\end{equation}
This corresponds to a plane wave propagating to the right. It may appear that
this class of solutions has 
two arbitrary functions of $u$, but one of them can be absorbed by 
reparametrizing $u$. The remaining free function can be taken to be the
amplitude of the dilaton field $\varphi$. There are
no independent gravitational degrees of freedom since we have assumed
isotropy in the transverse $x^i$ space. (Gravitational plane waves would
expand some directions while contracting others.) There are clearly analogous
solutions depending only on $v$ representing plane waves moving to the left.

We now construct a solution of the full equations (\ref{eins}) and 
(\ref{scalar}) by
matching a solution depending only on $u$ (for $y>0$) to a solution depending
only on $v$ (for $y<0$). The result is a solution in which the bulk spacetime
consists of plane waves moving away from the brane on both sides. We will call
this the {\it outgoing brane wave model}. 

The matching conditions at $y=0$
can be solved as follows. It will be convenient to
write
\begin{equation}
B(u,v)=\log h(u)  \label{+sideB}
\end{equation}
on the $y>0$ side. The continuity of $B$ implies that $B(u,v)=\log h(v)$ on
the $y<0$ side. Since, according to (\ref{b}), 
the jump in $\vp_{,y}$ must be proportional to 
the jump in $B_{,y}$ for all $t$, $\vp$ itself must be proportional to $B$.
\begin{equation}
\varphi (u,v)=-\frac{9}{4}\,b\,\log h(u)  \label{+sidephi}
\end{equation}
for $y>0$ and $\varphi (u,v)=-\frac{9}{4}b\log h(v)$ for $y<0$.
Note that an additive constant in $\varphi (u,v)$ can be absorbed by
scaling $h$ and the consequent additive constant in $B$ can be absorbed by
scaling $x^i$. We can now immediately integrate (\ref{Aeq}) to 
obtain\footnote{Throughout this paper it is understood that
all functions inside the logarithm have absolute value signs.}
\begin{eqnarray}
A(u,v) &=&\frac{1}{2}\left(1+\frac{9}{4}b^{2}\right)B(u,v)
+\frac{1}{2}\log B_{,u}(u,v)
\nonumber \\
&=&\frac{9}{8}\,b^{2}\,\log h(u)+\frac{1}{2}\log h^{\prime }(u)
\label{+sideA} 
\end{eqnarray}
for $y>0$ and $A(u,v)=\frac{9}{8}b^{2}\log h(v)+\frac{1}{2}\log
h^{\prime }(v)$ for $y<0$. An additive constant can be absorbed by
scaling $u$ and $v.$

The function $h$ is determined once one picks an equation of state for the
matter on the brane. In other words, the equation of states fixes both
the amplitude of the bulk waves, and the dynamics of the brane geometry.
Let
\be
P_T = \gamma\, \rho_T
\ee
where $\gamma$ is a constant, and $P_T$ and $\rho_T$ are the total pressure
and energy density. (A cosmological constant corresponds to $\gamma = -1$.)
Then the matching conditions (\ref{Bmatch}) and
(\ref{Amatch}) imply that the jump in $A_{,y}$ is proportional to the
jump in $B_{,y}$. Since this must hold for all time, this implies
\be 
A=-(2+3\gamma)B + k \label{AeqB}
\label{defk}
\ee
for some constant $k$.
This yields a first order equation for $h$ which can be integrated explicitly.

To illustrate some of the features of this model we start  with
a particularly simple special case. 
Inspection of (\ref{Aeq}) shows that we can choose $\varphi $, $A,$ and $B\ $%
to be linear functions of $u$ on the $y>0$ side and linear functions of $v$
on the $y<0$ side respectively. This corresponds to setting
\begin{equation}
\label{h1}
h(t)=e^{\lambda\, t}
\end{equation}
where we will assume the constant $\lambda$ is positive.
Thus, on the $y>0$ side, $B=\lambda\, u,$ 
$\varphi =-\frac{9}{4}\,b\,\lambda\, u,$
and 
\be
A=\frac{1}{2}\,(1+\frac{9}{4}\,b^{2})\,\lambda\, u + \frac{1}{2}\log \lambda 
\ee
and similarly on the $y<0$
side. We now set $b=\pm \frac{2}{3}$. (The case of general $b$ will be
considered below.)  Then $A=B $ + const, and it follows from (\ref{AeqB})
that $\gamma=-1$ so the
stress energy on the brane is a pure cosmological constant.
From (\ref{Bmatch}), the vacuum energy is 
\begin{equation}
V = 12\,\lambda^{\frac12}
\end{equation}

The bulk metric is
\begin{equation}
ds^{2}=e^{2\lambda (t-y)}[-\lambda dt^{2}+\lambda dy^{2}+dx_i dx^i]
\end{equation}
for $y>0$ and
\begin{equation}
ds^{2}=e^{2\lambda (t+y)}[-\lambda dt^{2}+\lambda dy^{2}+dx_i dx^i]
\end{equation}
for $y<0.$ 
Changing to cosmological time $\tau =e^{\lambda t}/\sqrt \lambda$, we
see that the metric on the brane at $y=0$ has the Robertson-Walker form $%
ds^{2}=-d\tau ^{2}+\lambda\tau ^{2}dx_i dx^i$ which after scaling $%
x^i$ gives
\begin{equation}
ds^{2}=-d\tau ^{2}+\tau ^{2}dx_i dx^i  \label{4d metric}
\end{equation}
Thus, with a constant vacuum energy $V$, we have a universe with the scale
factor growing linearly $a(\tau )=\tau $ rather the usual exponential growth.
It is important to note that the expansion rate is independent of the
value of the vacuum energy (as long as it is nonzero). 
Thus this solution has the self-tuning
feature described in \cite{KSS,HDKS}.

Unlike the static solutions discussed in \cite{KSS,HDKS}, this solution has
no naked timelike singularity. However there is a null
singularity at $u=-\infty$ on the right of the brane and at $v=-\infty$ on
the left. Null geodesics from these singularities reach the brane
in finite affine parameter
so they are not really at infinity. This is most easily seen by introducing
a new coordinate $U = e^{2\lambda u}/2$ so the metric for $y>0$
becomes
\begin{equation}
ds^{2}= -dUdv + 2 U \ dx_i dx^i
\label{rtmet}
\end{equation}
The singularity is now at $U=0$. The brane is located at $u=v$ or $U = 
e^{2\lambda v}/2$,  so the brane never actually hits the singularity, but
instead becomes asymptotically null as $v\rightarrow -\infty$ (see Fig.~1).
The geometry on the left of the brane is similar with the roles of 
$u$ and $v$ interchanged. 

\begin{figure}[t]
\centerline{\epsfysize=6truecm  \epsfbox{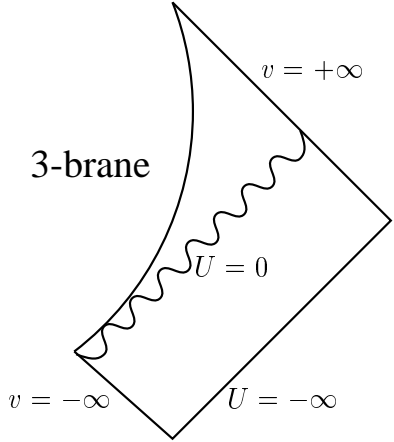} }
\vspace{1cm}
\tighten{
\caption[]{\it 
The causal structure on the right of the brane for the case 
$b=\pm\frac{2}{3}$. The brane never actually hits the
singularity in the bulk. }}
\end{figure}



Since the null singularity cuts off space in the fifth direction, 
one has a nonzero effective four dimensional Newton's constant $G_4$. But 
since the distance of the singularity 
to the brane changes with time, $G_4$ will be time
dependent. This is the main drawback of the outgoing  brane wave model.
It can perhaps be cured by modifying the bulk Lagrangian to 
try to localize gravity 
as in the Randall-Sundrum scenario.

As discussed in \cite{KSS}, the value of $b$ that we have chosen,
$b=\frac{2}{3}$, arises naturally in string theory since then (\ref{action})
is just
the Einstein frame action arising from a string frame source proportional to 
$e^{-2\vp}$. For our solution, the corresponding string metric $\tilde G_{MN}
=e^{4\vp/3} G_{MN}$ is flat! The bulk solution is simply a linear dilaton
vacuum where the dilaton is proportional to a null coordinate. This is well
known to be an exact solution to string theory \cite{MYERS}. The brane
geometry is also flat in the string frame.

The general solution for $h$ given an arbitrary dilaton coupling to the brane
$b$ and arbitrary equation of state $P_T=\gamma \rho_T$ is easy to construct.
From (\ref{+sideA}) and (\ref{AeqB}) we get a first order equation for $h$
with solution
\be
h(u) = (\lambda u)^{4/(20+24\gamma + 9b^2)}
\label{genhsol}
\ee
for $y>0$ where $\lambda$ is a positive constant related to $k$ in
(\ref{defk}).
The solution for $y<0$
is identical with $u$ replaced by $v$. This is valid whenever the
exponent is finite. For the special case considered above ($\gamma =-1$
and $b=\pm 2/3$) the exponent diverges and the solution for $h$
is not a power law, but rather an exponential as
we saw. The metric on the brane turns out to be
\be
ds^2 = -d\tau^2 + \tau^{8/[12(1+\gamma)+9b^2]}\ dx_i dx^i
\ee
after rescaling the $x_i$. In particular, for a pure cosmological constant
($\gamma=-1$) the brane geometry is 
\be
ds^2 = -d\tau^2 + \tau^{8/(9\,b^2)} dx_i dx^i
\label{purel}
\ee
Thus, for any nonzero coupling $b$, the expansion is a power law which is
independent of the value of the vacuum energy.
The value of this vacuum energy can be computed from (\ref{Bmatch}).
When there is no other matter on the brane ($\gamma=-1$) 
one finds that
the vacuum energy is indeed constant on the brane and given by
\be
V= \pm {24\over |9\,b^2-4|^\frac12}\; \lambda^{\frac12}.
\ee
where the sign is the same as the sign of $9b^2 -4$.
The
causal structure of the resulting spacetime is shown in Fig.~2. In this case,
the singularity is at $u=0$ and the brane hits the singularity at the time
of the big bang. 
\begin{figure}[t]
\centerline{\epsfysize=7truecm  \epsfbox{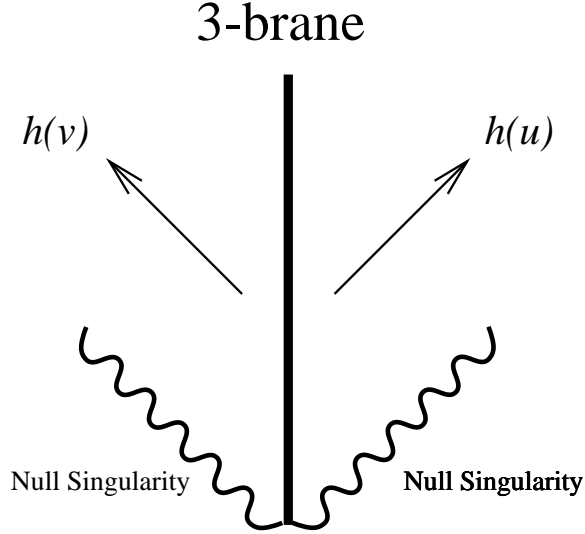} }
\vspace{1cm}
\tighten{
\caption[]{\it 
The causal structure of the outgoing brane wave model }}
\end{figure}

The properties of
this solution are similar to the case $b=\frac{2}{3}$ discussed above.
(In fact, Fig. 2 accurately represents the causal structure of the
complete spacetime in this case also.)
It is even true that the bulk spacetime is still
an exact solution to string theory.
In terms of a new null coordinate $U$, the bulk metric becomes
\be
ds^2= - dUdv + U^{8/(4+9b^2)} dx_idx^i
\label{genmet}
\ee
Surprisingly, this metric is independent of the equation of state parameter
$\gamma$. However, the position of the brane in $U,v$ coordinates now
depends on $\gamma$.
The metric (\ref{genmet}) has a covariantly constant
null Killing vector $ \ell = \p/\p v$ in addition to the translational
symmetries in $x^i$. So it belongs to a general class 
of metrics
known as exact plane waves. The Riemann tensor involves two powers of $\ell$
so all scalar curvature invariants vanish. Nevertheless, the singularity
is real and results in infinite tidal forces. From the standpoint of 
string theory, since the curvature is null,
all $\alpha'$ corrections automatically vanish \cite{HS}. 
Moreover, one can show that the exact two dimensional beta functions
of a sigma-model with this target space vanish,
so this is an exact solution to string theory even nonperturbatively \cite{Ama}.

When
$b=0$ (and $\gamma=-1$) (\ref{genhsol}) shows that $h = (\lambda u)^{-1} $. The
bulk metric is then $ds^2 = (\lambda u)^{-2} (-\lambda dudv + dx_i dx^i)$
which is actually flat space in disguise. The metric on the brane
is $-d\tau^2 + e^{-2\tau\sqrt \lambda} dx_i dx^i$ and we recover the usual
de Sitter metric. 

We now ask what happens
if there is a transition from one value of the vacuum energy $V_{1}$
to another value $
V_{2}.$ Since a pure vacuum energy is always constant, to achieve this
we must assume that the matter on the brane consists
of both a vacuum energy and another component which we take to be radiation.
Physically, some of the initial vacuum energy is converted to radiation
which is then redshifted away leaving a smaller vacuum energy in the future.
Since the two independent components of the stress energy tensor (\ref{Bmatch})
 and (\ref{Amatch})
are determined in terms of $A,B$ and $\vp$ and these three functions are
determined in terms of $h$ via (\ref{+sideB}), (\ref{+sidephi}), and
(\ref{+sideA}), we can compute $V(t)$ and $\rho(t)$ given any $h(t)$.
Setting $P=\rho/3$ we find
\begin{equation}
V=\frac{3\,h^{\frac{9}{8}b^{2}}}{|h^{\prime }|^{\frac{1}{2}}}
\left[\left(3+\frac{9}{8}b^{2}\right)\frac{h^{\prime }}{h}+
\frac{h^{\prime \prime }}{2h^{\prime }}\right]
\end{equation}
and
\begin{equation}
\rho =\frac{3\,h^{\frac{9}{8}b^{2}}}{|h^{\prime }|^{\frac{1}{2}}}
\left[\left(1-\frac{9}{8}b^{2}\right)\frac{h^{\prime }}{h}-\frac{h^{\prime
\prime }}{2h^{\prime }}\right]
\label{rho}
\end{equation}
Here everything is evaluated on the brane so that $h$ stands for 
$h(t),$ $h^{\prime }$ for $\frac{dh(t)}{dt},$ and so forth.

To show that  the brane geometry remains independent
of the value of the vacuum energy even after a phase transition,
one must find a function $h(t)$ for which $\rho\ge 0 $ everywhere, $\rho
\rightarrow 0$ in the past and future, and $V$ approaches different constants
in the past and future.
Since (\ref{purel}) is the only solution for pure vacuum energy,
the geometry will expand at this rate 
both in the past and future. The only difference will be a constant 
rescaling of the $x^i$ coordinates. It is not hard to find a suitable
function $h(t)$. For example,
in the case $b=\pm \frac{2}{3}$, we can choose
\be
h(t)=\frac{e^{\lambda _{1}t}}{
1+e^{(\lambda _{1}-\lambda _{2})t}}
\label{sampleh}
\ee
with $\lambda _{1}>\lambda _{2}$.
This describes a universe with pure vacuum energy $V_1= 12 \lambda_1^{1/2}$
in the past and $V_2= 12 \lambda_2^{1/2}$ in the future.
During intermediate times there is radiation with energy density given
by substituting (\ref{sampleh}) into (\ref{rho}).
In Fig.~3 we
show the time dependence of $V(t)$ and $\rho(t)$, assuming radiation only,
for the case $\lambda_1=2.5$, $\lambda_2=1$ and $b=2/3$.
\begin{figure}[t]
\label{vr}
\centerline{\epsfysize=10truecm  \epsfbox{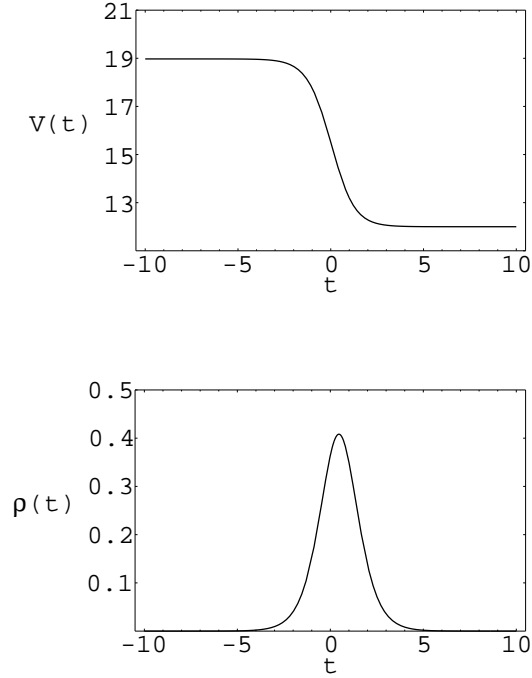} }
\tighten{
\caption[]{\it Time dependence of $V(t)$ and $\rho(t)$ using
$\lambda_1=2.5$ and $\lambda_2=1$ in {\rm (\ref{sampleh})}
with $b=2/3$. We assume radiation
only and thus $P=\rho/3$.}}
\end{figure}

\sect{General Time Dependent Solutions}
\label{gen}

We now turn to more general time dependent solutions of the bulk equations
of motion (\ref{153}) -(\ref{ii15}).
Remarkably, one can find explicit solutions which are much more general than
the plane waves discussed so far. One can then match these solutions on 
the brane to obtain solutions to the full field equations.

The key is to observe that (\ref{153}) involves
only $B$ and can be solved to give
\begin{equation}
B(u,v)=\frac{1}{3}\log (f(u)+g(v))  \label{B}
\end{equation}
where $f$ and $g$ are two arbitrary functions. (A possible additive integration
constant can always be absorbed by an overall scaling of $f$ and $g$.)

Given $B(u,v),$ we can solve the dilaton equation of motion (%
\ref{dil3}). Using separation of variables, the general solution 
can be written

\begin{equation}
\varphi (u,v)=\int dk\frac{c(k)}{\sqrt{(f(u)-k)(g(v)+k)}} \label{vpsol}
\end{equation}
 A special case is obviously
\begin{equation}
\varphi (u,v)=\frac{c}{\sqrt{(f(u)-k)(g(v)+k)}}
\end{equation}
For this case, we can solve (\ref{uu}) and (\ref{553}) explicitly
to find $A(u,v)$
with the result
\begin{equation}
A(u,v)=-\frac{1}{3}\log (f(u)+g(v))+\frac{1}{2}\log (f^{\prime }(u)g^{\prime
}(v))-\frac{c^{2}}{12}\frac{(f(u)+g(v))^{2}}{((f(u)-k)(g(v)+k))^{2}}+a
\end{equation}

Another special solution for $\vp$ which will be more useful in the following 
is
\begin{equation}
\label{spe1}
\varphi (u,v)=\frac{\alpha }{3}\log (f(u)+g(v))
\end{equation}
and the associated solution for $A$ is
\begin{equation}
\label{spe2}
A(u,v)=\left(\frac{2\alpha ^{2}}{9}-1\right)\frac{1}{3}\log (f(u)+g(v))
+\frac{1}{2}\log
(f^{\prime }(u)g^{\prime }(v))+a  \label{A}
\end{equation}
with $\alpha $  and $a$ integration constants.

As an aside we mention one other special solution 
\begin{equation}
\varphi (u,v)=f(u)-g(v)
\end{equation}
and
\begin{equation}
A(u,v)=-\frac{1}{3}\log (f(u)+g(v))+\frac{1}{2}\log (f^{\prime }(u)g^{\prime
}(v))+\frac{1}{3}(f(u)+g(v))^{2}+a
\end{equation}
This solution is not interesting for our purposes since the matching 
conditions
(\ref{Bmatch}), (\ref{Amatch}), and (\ref{phimatch})
can only be satisfied if the stress-energy tensor on the brane 
vanishes\footnote{Continuity of $\varphi $ and $B$ 
requires $f_{+}=f_{-}$ and $g_{+}=g_{-}$ (where the subscripts denote the
solution for $y>0$ or $y<0$).
 From 
 (\ref{Bmatch}) - (\ref{phimatch}) we see that this leads 
to the conclusion
that on the brane $\rho_{T}=P_{T}=0.$}.

\sect{Matching to an initially static solution}

We can now address the question of what happens if one starts with
the static solutions discussed in \cite{KSS,HDKS} 
and then changes the vacuum energy.
We begin by 
writing down the static solution with our form of metric. For our solutions 
to be independent of $t$, the functions $A,$ $B,$
and $\varphi $ can depend on  $(v-u)$ only. Since we want the metric on  the
brane to be Poincare invariant, we set $A=B$. With these restrictions,
the bulk equations (\ref{153}) -(\ref{ii15})
reduce to two equations:
\begin{eqnarray}
A^{\prime \prime }(y)+3A^{\prime }(y)^{2} &=& 0, \\
9A^{\prime }(y)^{2}-\varphi ^{\prime }(y)^{2}&=& 0,
\label{statvp}
\end{eqnarray}
We wish to solve these equations for $y>0$ and $y<0$ separately and then
match at $y=0$. Denoting the solutions in the two regions by subscripts $+$
or $-$, we have for $y>0$
\begin{equation}
A(y)=\frac{1}{3}\log (y_{+}-y)+a_{+},  \label{apeq}
\end{equation}
and for $y<0$
\begin{equation}
A(y)=\frac{1}{3}\log (y+y_{-})+a_{-}  \label{ameq}
\end{equation}
where $y_\pm$ and $a_\pm$ are constants of integration. In order
to have a finite
4D Planck mass  \cite{KSS,HDKS} are forced to have naked singularities
at $y=y_+$ and $y=-y_-$ to cut off spacetime.
So the relevant region of spacetime is $-y_- < y < y_+$.
Making the appropriate coordinate transformation we see that these agree
with the solutions given in \cite{KSS} which we will refer to as KSS. From the
dilaton equation (\ref{statvp}), we have
the freedom to choose on either side of the brane $\varphi =+3A$ or $\varphi
=-3A$ up to a constant. Let us fix
\begin{equation}
\varphi (y)=\log (y_{+}-y)+d_{+}
\end{equation}
for $y_{+}>y>0.$ In the nomenclature of KSS, for $0>y>-$ $y_{-}$ the choice
\begin{equation}
\varphi (y)=-\log (y+y_{-})+d_{-}
\end{equation}
is known as a type I solution, while the choice
\begin{equation}
\varphi (y)=\log (y+y_{-})+d_{-}
\end{equation}
is known as type II. 
The continuity of $\varphi $ and $A$ at $y=0$
determines $d_{+}$ and $a_{+}$ in terms of the other constants. 

Since we have set $A=B$, the matching conditions (\ref{Bmatch}) and
(\ref{Amatch}) imply that the stress-energy tensor on the brane takes the 
form of a pure vacuum energy density, with
\begin{equation}
\left.-6\,A_{,y}\right]
=2\left( \frac{1}{y_{+}}+\frac{1}{y_{-}}\right) =\left.e^{A+b\varphi
}\right|V  \label{ajump}
\end{equation}
The matching condition for $\vp$ (\ref{phimatch}) implies
\begin{equation}
\frac{8}{3}\,\left. \varphi_{,y} \right]
=-\frac{8}{3}\left( \frac{1}{%
y_{+}}-\frac{1}{y_{-}}\right) =\left.b\,e^{A+b\varphi }\right|V  \label{phjump}
\end{equation}
for type I and
\begin{equation}
\frac{8}{3}\,\left. \varphi_{,y} \right]
=-\frac{8}{3}\left( \frac{1}{%
y_{+}}+\frac{1}{y_{-}}\right) =\left.b\,e^{A+b\varphi }\right| V
\end{equation}
for type II. Taking the ratio of $\left.\varphi^{\prime}\right]$ and
$\left.A^{\prime }\right]$ we see that
type I is allowed only if $b\neq \pm \frac{4}{3}$ (assuming that both
$y_{+}$ and $y_{-}$ are finite) while
type II is allowed only if $b=-\frac{4}{3}$. The solution for $b=+\frac{4}{3}$
is obtained from the type II solution by simply 
changing the sign of $\vp$ on both sides of the brane
so $\vp =-3A$ everywhere.

Suppose now that there is a phase transition in the microphysics so
that $V$ changes from one constant to another. Will the brane remain flat?
We found
that it is most easy to use the special bulk solution (\ref{spe1}) and
(\ref{spe2}) to
 match onto the type II solution in KSS.
By causality, the static solution must persist
for the region of the bulk defined by $u=t-y<0$ and $v=t+y<0$. 
We thus take the solution (\ref{B}) for $B$ to be the following form
\begin{equation}
\label{statb1}
B(u,v)=\frac13\log(y_*(u)+u-v)
\end{equation}
for $y>0$ and, for $y<0$,
\begin{equation}
\label{statb2}
B(u,v)=\frac13\log(y_*(v)+v-u),
\end{equation}
In writing down (\ref{statb1})
and (\ref{statb2}) we have used 
the continuity condition of $B(u,v)$ at the location
of the 3-brane $u=v$. $y_*$ describes the time evolution of the
location of the naked singularity on both sides. Before the phase 
transition, the universe is static and we have $y_*=$constant. Given 
$B(u,v)$ we obtain $\varphi(u,v)$ and $A(u,v)$ for $y>0$
\begin{eqnarray}
\label{eps1}
\varphi(u,v)&=& \epsilon \log(y_*(u)+u-v) + d, \\
A(u,v) &=& \frac13\log(y_*(u)+u-v)+ \frac12\log(1+y_*^{\prime}(u))+a
\end{eqnarray}
and for $y<0$
\begin{eqnarray}
\label{eps2}
\varphi(u,v)&=& \epsilon \log(y_*(v)+v-u) + d, \\
A(u,v) &=& \frac13\log(y_*(v)+v-u)+ \frac12\log(1+y_*^{\prime}(v))+a.
\end{eqnarray}
$d$ and $a$ are integration constants and $\epsilon=\pm1$. 
This solution was determined as follows.
Demanding Poincare invariance on the brane for the initially static solution
sets $A=B$ and thus
$\alpha^2=9$ in (\ref{spe2}). The continuity condition
for $\varphi(u,v)$ requires $\epsilon$ to take the same sign on both
sides of the brane. (\ref{b}) determines $b=-\epsilon \frac43$.
It is clear that, for $y_*=$constant$>0$, this solution
reduces to the type II static solution discussed in KSS. 

Using the matching conditions (\ref{Bmatch}) and (\ref{Amatch})
we obtain
\begin{eqnarray}
\label{rhostat}
\rho_T(t) &=& 4 \, e^{-a- b\,d}\,
  \frac{y_*^{\prime}(t) +2}{\left|y_*^{\prime}(t)+1\right|^{\frac12}}, \\
\label{rpstat}
\rho_T(t)+P_T(t) &=& -2 \, e^{-a- b\,d}\,
  \frac{y_*(t)\;y_*^{\prime\prime}(t)}
{\left|y_*^{\prime}(t)+1\right|^{\frac12}\;(y_*^{\prime}(t)+1)}.
\end{eqnarray}
From (\ref{rhostat}) we see immediately that $\rho_T$ depends 
on $y_*^{\prime}$ only, but not on the 
value of $y_*$. This means that, in order for the vacuum energy density
$V$ to settle to a constant value after the phase transition, 
$y_*^{\prime}(t)$ has to be a constant in the future. 

There are two possible scenarios following from this observation. 
The first is that $y_*(t)$ settles to a constant value 
and thus $y_*^{\prime}=0$. (\ref{rhostat}) and (\ref{rpstat}) implies
the vacuum energy density $V$ would always go back to its original
value before the phase transition and 
the universe becomes static again.
In this scenario the vacuum energy density is determined by the constants
of integration $a$ and $d$ and is independent of the value of
$y_*$.

The second scenario is that $y_*(t)$ becomes linear in $t$ for $t>0$. In
this case, the vacuum energy density $V$ could change to another constant
value. However, the universe is not static even after the phase transition
and would start evolving thereafter. As an example we take 
$y_*(t)=y_0 - \xi\,t$ with $0<\xi<1$, then the vacuum energy density changes to
\begin{eqnarray}
V_2 &=& 4\, e^{-a-b\,d} {2-\xi \over \sqrt{1-\xi}} \nonumber\\
    &=& {V_1\over 2}\;{2-\xi \over\sqrt{1-\xi}}, 
\end{eqnarray}
where $V_1$ is the initial value of the vacuum energy density. One can check
that the transition can occur with positive energy density in the radiation.
After the transition
the universe is no longer static, but evolves toward a `big crunch' 
singularity.

Here we have constructed an example which matches onto the initially static
type II solution of KSS by using the special bulk solution (\ref{spe1}) and
(\ref{spe2}). The continuity condition of $\varphi$ requires
$\epsilon$ to take the same sign in (\ref{eps1}) and (\ref{eps2}).
 We are thus unable to match onto the type I solution of KSS using 
this special solution since the type I solution needs $\epsilon$ to 
take opposite sign. It would be interesting to try to use the general 
solutions in Section \ref{gen} to match onto the type I solution
of KSS and study scenarios with phase transitions.

\acknowledgments
We thank Lisa Randall,  Eva Silverstein and Erik Verlinde
for informative discussions.
We also would like to thank Daniel Holz for help making the figures. 
 This work was supported in part by the
NSF under grant numbers PHY89-04035 and PHY95-07065, and also
by Department of Energy under
grant number DOE-ER-40682-143. I.\,L. is supported in part by an ITP
Graduate Fellowship.

\tighten


\end{document}